\documentclass[a4paper,11pt]{article}
\usepackage{jheppub} 
\usepackage{lineno}

\usepackage{graphicx,subfigure}
\usepackage{amsmath}
\usepackage{amssymb}
\usepackage{physics}
\usepackage{xcolor}
\usepackage{slashed,cancel}
\usepackage{multirow}
\usepackage{float}
\usepackage{capt-of}
\usepackage{comment}
\usepackage{ mathrsfs }
\usepackage{enumitem}
\usepackage{rotating}
\usepackage[colorlinks=true,linkcolor=blue,urlcolor=blue,citecolor=blue]{hyperref}

\arxivnumber{2411.19171} 

\title{\boldmath Looking into the quantum entanglement in $H\to ZZ^\star$ at LHC within SMEFT framework}

\author[a]{Amir Subba \note[a]{Corresponding Author}}
\author[b]{,Ritesh K. Singh }
\author[c]{and Rohini M. Godbole \note[c]{Deceased}}
\affiliation[a,b]{Department of Physical Sciences, Indian Institute of Science Education and Research Kolkata,\\ Mohanpur, 741246, India}
\affiliation[c]{Center of High Energy Physics, Indian Institute of Science,\\ Bengaluru 560012, India}

\emailAdd{as19rs008@iiserkol.ac.in, ritesh.singh@iiserkol.ac.in}

\abstract{We study $H\to ZZ^\star$ production process in final four lepton states at $13$ TeV LHC in SMEFT framework. The anomalous $HZZ$ couplings are parameterized with dimension-6 $SU(2)_L\times U(1)_Y$ gauge invariant operators. We compute the eight polarizations of each $Z$ boson and $64$ spin-correlations as asymmetries in angular functions of final decayed leptons in the rest frame of the $Z$ boson. These asymmetries are further used to construct the joint density matrix (DM) for $ZZ^\star$ system. However, such DM suffers from negative probability and eigenvalues. To alleviate the negativity issues, we reconstruct the DM using asymmetries of symmetrized angular functions owing to the indistinguishability of two $Z$ bosons. The symmetrized DM is further employed to compute lower bound for concurrence as a witness of entanglement measurable at the collider experiments. The $ZZ^\star$ system is found to be in an entangled state for all values of the anomalous couplings. Notably, while the lower bound exhibits poorer sensitivity to anomalous couplings compared to asymmetries, it demonstrates distinct behavior for CP-even and odd couplings.}

\begin{document}
\maketitle
\flushbottom

\section{Introduction}
\label{sec:intro}

Entanglement is a quantum signature of a system with no analog to classical correlations. Starting with a classic paper~\cite{PhysRev.47.777} by Einstein, Rosen, and Podolsky, which questioned the incompleteness of quantum theory to explain causality and locality, entanglement today has a central place in quantum computation~\cite{Shor:1995hbe,Steane:1996ghp}, quantum teleportation~\cite{Bennett:1992tv}, quantum dense coding~\cite{Bennett:1992zzb}, and quantum cryptography~\cite{Ekert:1991zz,Jennewein:2000zz}. Though the measure of entanglement was extensively studied at a low energy scale~\cite{PhysRevLett.49.91,Hensen:2015ccp,PhysRevLett.115.250401}, there is a recent surge in literature exploring the high energy regime to quantify the entanglement. The theoretical exploration of entanglement has recently been extended to collider settings involving a variety of fundamental particles including quarks, vector boson, and Higgs boson~\cite{Barr:2022wyq,Aguilar-Saavedra:2022mpg,Altakach:2022ywa,Cheng:2023qmz,Han:2023fci,Dong:2023xiw,Varma:2023gwh,Barr:2024djo,Fabbrichesi:2023cev,Subba:2024mnl,Morales:2023gow,Morales:2024jhj,Fabbrichesi:2023jep,Grossi:2024jae,Aoude:2023hxv,Marzola:2023oyv}, as well as previously between taus produced from the $Z$ boson decay~\cite{Abel:1992kz}. Recent paper by ATLAS~\cite{ATLAS:2023fsd} and CMS~\cite{CMS:2024pts} collaboration of LHC reports the existence of entanglement in top quark pair. Measurement of entanglement in a two-qubit case is done with the parameters of the density matrix, the parameters being polarization and spin correlations calculated from the angular distribution of final decayed fermions. It has been shown~\cite{Aguilar-Saavedra:2022uye} that the measurement of bipartite qubit entanglement can be achieved with just the spin correlation matrix. The measure of entanglement is also studied in the beyond the SM (BSM) scenarios~\cite{Aoude:2023hxv,Aoude:2022imd,Aoude:2023hxv,Severi:2022qjy,Bernal:2023ruk,Bernal:2024xhm,Sullivan:2024wzl,Fabbrichesi:2023jep}.
	 \\
	The dynamics of nature in different energy scales can be explained by different subsets of theories when we lack the complete structure of the fundamental governing theory. In the absence of a definitive renormalizable theory for BSM physics, the effects of potential new physics at high energies $\Lambda \gg v$ can be systematically parameterized in a model-independent manner using higher-dimensional operators. This framework, known as the Standard Model Effective Field Theory~(SMEFT)~\cite{Weinberg:1978kz,Weinberg:1980wa,Buchmuller:1985jz}, expands the SM to incorporate these effects at the electroweak scale. Various experimental measurements of observables at the electroweak scale can then be used to constrain or determine the coefficients of these higher-dimension operators.
	\\
	In this article, we investigate the \(H \to ZZ^\star\) process at the 13 TeV LHC, focusing on events with four-lepton final states in the presence of anomalous \(HZZ\) couplings at leading order. Our primary objective is to study the entanglement properties of the \(ZZ^\star\) system, employing quantum tomography through angular functions of the final-state leptons. A comprehensive and realistic analysis, however, necessitates the inclusion of higher-order quantum corrections and fiducial cuts, which can influence the helicity structure of the production mechanism~\cite{Grossi:2024jae}. However, in this work, we emphasize challenges in reconstructing the density matrix (DM) for systems involving two identical bosons at the leading order. We also propose a refined parameterization of the DM to address these issues, which will be elaborated upon in a later section. Furthermore, we explore the behavior of the entanglement witness under different CP structures of potential new physics. Several studies on entanglement in the \(H \to ZZ^\star\) process are available in the literature~\cite{Aguilar-Saavedra:2022wam,Sullivan:2024wzl,Ruzi:2024cbt,Bernal:2024xhm,Aguilar-Saavedra:2024whi,Bernal:2023ruk}, which provide additional context and insights.	
	\\
	At the leading order, single Higgs production via gluon-gluon fusion happens through the quark loop, which one can parameterize to effective couplings at large top quark mass limit~($m_t > m_H$). The effective Lagrangian relevant to the coupling of Higgs boson with two gluons is,
	\begin{equation}
		\mathcal{L}_{ggH} = -\frac{1}{4}g_H G^a_{\mu\nu}G^{a,\mu\nu}H,
	\end{equation}
	where the field tensor is given as $G^a_{\mu\nu} = \left(\partial_\mu G_\nu - \partial_\nu G_\mu - g_s f^{abc}G^b_\mu G^c_\nu\right)$ and the effective coupling along with the loop coefficient is defined as~\cite{Kauffman:1996ix,Georgi:1977gs}
	\begin{equation}
		\begin{aligned}
			g_H &= \frac{g_s^2}{12\pi v}\left(1+\frac{7}{30}\left(\frac{m_H}{m_t}\right)^2+\frac{2}{21}\left(\frac{m_H}{m_t}\right)^4+\frac{26}{525}\left(\frac{m_H}{m_t}\right)^6\right). 
		\end{aligned}
	\end{equation}
	The general Higgs boson coupling with two weak $Z$ boson involving both CP-even and odd couplings is~\cite{Bernal:2023ruk,Zagoskin:2015sca,Godbole:2007cn},
	\begin{equation}
		\label{eqn:HZZ}
		\begin{aligned}
			\Gamma^{HZZ}_{\mu\nu} &= \frac{igm_Z}{\cos\theta_W}\left[v_1\eta_{\mu\nu} + v_2(k+p)_\mu(k+p)_\nu + v_3\epsilon_{\alpha\beta\mu\nu}(k+p)^\alpha(k-p)^\beta\right],
		\end{aligned}
	\end{equation}
	where $v_1,v_2$ are CP-even parameters and $v_3$ is CP-odd. At the SM tree level, $v_1=1, v_2=0=v_3$. The Lagrangian that could generate the above given vertex can be~\cite{Bolognesi:2012mm,Godbole:2007cn}, \begin{equation}
		\label{eqn:lagvertex}
		\begin{aligned}
			\mathcal{L} &\sim v_1 HZ_\mu Z^\mu + v_2\left(HZ_{\mu\nu}Z^{\mu\nu} + Z_{\mu\alpha}Z^{\nu\beta}\left[\partial_\beta\partial_\alpha H\right] + v_3 H Z_{\mu\nu}\widetilde{Z}^{\mu\nu}\right),
		\end{aligned}
	\end{equation}
	where $Z_{\mu\nu} = \partial_\mu Z_\nu - \partial_\nu Z_\mu$, and the dual field $\widetilde{Z}_{\mu\nu}$ is defined as $\widetilde{Z}_{\mu\nu}=1/2\epsilon_{\mu\nu\alpha\beta}Z^{\alpha\beta}$, with Levi-Civita tensor following a standard notation, i.e., $\epsilon_{0123}=1$. The anomalous contribution to the $HZZ$ couplings can be parameterized with higher order $SU(2)_L\times U(1)_Y$ gauge invariant operators. In this work, we limit ourselves to dimension-6 operators and the relevant operators inducing anomalous $HZZ$ couplings in HISZ basis~\cite{Hagiwara:1993ck} are
    \begin{equation}
		\label{eqn:lagdim6}
		\begin{aligned}
			&\mathscr{O}_{W} &=& ~~\left(D_\mu\Phi\right)^\dagger W^{\mu\nu}\left(D_\nu\Phi\right),\\
			&\mathscr{O}_B&=&~~\left(D_\mu\Phi\right)^\dagger B^{\mu\nu}\left(D_\nu\Phi\right),\\
			&\mathscr{O}_{WW} &=&~~ \Phi^\dagger W_{\mu\nu}W^{\mu\nu}\Phi, \\ 
			&\mathscr{O}_{BB} &=&~~\Phi^\dagger B_{\mu\nu}B^{\mu\nu} \Phi,\\  &\mathscr{O}_{\widetilde{W}} &=&~~ \left(D_\mu\Phi\right)^\dagger \widetilde{W}^{\mu\nu}\left(D_\nu\Phi\right),\\
			&\mathscr{O}_{\widetilde{W}W} &=&~~ \Phi^\dagger W_{\mu\nu}\widetilde{W}^{\mu\nu}\Phi,\\
			&\mathscr{O}_{\widetilde{W}W} &=&~~
			\Phi^\dagger B_{\mu\nu}\widetilde{B}^{\mu\nu} \Phi. 
		\end{aligned}    
	\end{equation}
	In the above equation the covariant derivative and the field tensors are defined as,
	\begin{equation}
		\begin{aligned}
			&D_\mu\Phi &=&~~ \left(\partial_\mu + \frac{i}{2}g_1 B_\mu + ig_W\frac{\tau^a}{2}W^a_\mu\right)\Phi,\\
			&W_{\mu\nu} &=&~~ i\frac{g_W\sigma^I}{2}W^I_{\mu\nu},\\
			&W^I_{\mu\nu} &=&~~ \partial_\mu W^I_\nu - \partial_\nu W^I_\mu -g_W\epsilon^{IJK}W^J_\mu W^K_\nu,\\
			&B_{\mu\nu} &=&~~ i\frac{g_1}{2}\left(\partial_\mu B_\nu - \partial_\nu B_\mu\right).
		\end{aligned}
	\end{equation}
    The two operators $\mathscr{O}_{WW}$, and $\mathscr{O}_{BB}$ induce a change in the two point function of weak field, which requires field renormalization in order to make the Lagrangian canonical. We redefine the operators as,
	\begin{equation}
		\begin{aligned}
			&\mathscr{O}_{WW} &=&~~ \left(\Phi^\dagger\Phi -v^2/2\right)W_{\mu\nu}W^{\mu\nu},\\
			&\mathscr{O}_{BB} &=&~~ \left(\Phi^\dagger\Phi-v^2/2\right)B_{\mu\nu}B^{\mu\nu}.
		\end{aligned}
	\end{equation}
    This redefinition removes the anomalous two point function of weak boson. The effective $HZZ$ Lagrangian in presence of above listed operators can be written as
	\begin{equation}
		\begin{aligned}
			\delta\mathscr{L}^{(6)} &= g^{(1)}_{HZZ}Z_{\mu\nu}Z^\mu\partial^\nu H + g_{HZZ}^{(2)} H Z_{\mu\nu}Z^{\mu\nu} + \widetilde{g}_{HZZ}^{(1)}\widetilde{Z}_{\mu\nu}Z^\mu \partial^\nu H + \widetilde{g}_{HZZ}^{(2)} HZ_{\mu\nu}\widetilde{Z}^{\mu\nu}.
		\end{aligned}
	\end{equation}
	The associated vertex factor in terms of the Wilson coefficient associated with the dim-6 operators are
	\begin{equation}
		\begin{aligned}
			g_{HZZ}^{(1)} &= \left(\frac{g^2_Wv}{2\Lambda^2}\right)\frac{\cos^2\theta_Wc_W+\sin^2\theta_Wc_B}{2\cos^2\theta_W},\\
			g_{HZZ}^{(2)} &= -\left(\frac{g^2_Wv}{2\Lambda^2}\right)\frac{\sin^2\theta_Wc_{BB}+\cos^4\theta_Wc_{WW}}{2\cos^2\theta_W},\\
			\widetilde{g}_{HZZ}^{(1)} &= \left(\frac{g^2_Wv}{16\Lambda^2}\right)c_{\widetilde{W}},\\
			\widetilde{g}_{HZZ}^{(2)} &= -\frac{g^2_Wv\sin^2\theta_W\tan^2\theta_W}{4\Lambda^2}c_{\widetilde{W}W} + \frac{g^2_Wv\cos^2\theta_W}{8\Lambda^2}c_{\widetilde{B}B},
		\end{aligned}
	\end{equation}
	with $\Lambda$ as some characteristic new physics scale and $c_i^\prime$s are the Wilson coefficient (WC) associated with dim-6 operators. The effect of some heavy physics are encoded in these coefficient once the heavy states are integrated out. The above Lagrangian is implemented in a {\tt FeynRules}~\cite{Christensen:2008py,Alloul:2013bka} to obtain a publicly available Universal FeynRules model~(UFO)~\cite{Degrande:2011ua,Darme:2023jdn}. For event generation at the leading order, we employ {\tt MadGraph5$\_$aMC@NLO}~\cite{Alwall:2014hca}. 
    \\\\
    The paper is organized as follows: In Section~\ref{sec:asymm}, we discuss the polarization and spin correlations of the $ZZ^\star$ system and their relation to asymmetries in the angular functions of the final-state fermions. Additionally, we outline the reconstruction of the correlated density matrix using these asymmetries. Section~\ref{sec:ent} focuses on the measure of entanglement in the $ZZ^\star$ system in the presence of anomalous $HZZ$ couplings, highlighting how the lower bound for concurrence behaves with anomalous couplings. Finally, we present our conclusions in Section~\ref{sec:conclude}.

\section{Angular distribution and Density Matrix}
	\label{sec:asymm}
	\noindent
	The density matrix encodes the maximal information of the quantum state in question. Consider a general case where not all of $N$ systems of the ensemble are in the same state, i.e., $N_i$ systems are in the state $\ket{\psi_i}$. We can write the mixed state as a convex sum, i.e., a weighted sum with $\sum_i p_i = 1$, of pure state density matrices,
	\begin{equation}
		\rho_{\text{mix}} = \sum_i p_i \rho_i^{(pure)} = \sum_i p_i \ket{\psi_i}\bra{\psi_i},
	\end{equation} 
	where $p_i$ is the probability of finding an individual system of the ensemble described by the state $\ket{\psi_i}$. The density matrix for a multi-particle system is defined in the Hilbert space $\mathcal{H}_1\otimes \mathcal{H}_2 \otimes \dots \mathcal{H}_n$ which is of the dimension $d = (2j_1+1)(2j_2+1)\dots (2j_n+1)$, and which has $d^2-1$ real independent parameters. The task of quantum state tomography is to determine each density matrix parameter for a single or multi-particle density matrix from experimental data. Different parameterization exist to represent the density matrix based on different choices of operators. We will work here in the spin parameterization based on $d^2-1$ traceless Hermitian operators $J^{(d)}$ of the $SU(d)$ group. For $d=3$, they are given by eight $J_i$ matrices listed in the Appendix~\ref{sec:jbasis}. In the current work, we study the Higgs boson decay to two spin-1 $Z$ bosons, where the density matrix for a single $Z$ boson can be expressed as,
	\begin{equation}
		\rho_Z = \frac{1}{3}\mathbb{I}_3 + \sum_{i=1}^8 a_i J_i,
	\end{equation}
	where $\mathbb{I}_3$ is the $3\times 3$ identity matrix and $a_i^\prime$s are the real parameters which form a $8$-dimensional Bloch vector. For a bipartite system ($ZZ^\star$), the density matrix can be parameterized as,
\begin{equation}
		\begin{aligned}
			\rho_{ZZ^\star} &= \frac{1}{9}\mathbb{I}_3 \otimes \mathbb{I}_3 + \frac{1}{3}\sum_{i=1}^8 a_iJ_i \otimes \mathbb{I}_3 + \frac{1}{3}\sum_{i=1}^8 \mathbb{I}_3 \otimes b_i J_i + \sum_{i=1}^8\sum_{j=1}^8 c_{ij}J_i \otimes J_j.
		\end{aligned}
	\end{equation}
	The $a^\prime$s and $b^\prime$s are the eight polarizations of two $Z$ bosons, and the $c_{ij}$ are the correlation parameters of the joint $ZZ^\star$ system. We extract these $80$ parameters of density matrix from the joint angular distribution of final leptons in the center-of-mass frame of the $ZZ^\star$ system, and each $Z$ boson is Lorentz transformed to their rest frame. The joint angular distribution of final leptons is given by~\cite{Rahaman:2021fcz},
	\begin{equation}
		\label{eq:jointad}
		\begin{aligned}
			\frac{1}{\sigma}\frac{d^2\sigma}{d\Omega^{l_1}d\Omega^{l_2}} &= \frac{9}{16\pi^2}\sum_{\lambda^\prime s}\rho_{ZZ^\star}\left(\lambda_{Z_1},\lambda_{Z_1}^\prime,\lambda_{Z_2},\lambda_{Z_2}^\prime\right)\times \Gamma_{Z_1}\left(\lambda_{Z_1},\lambda_{Z_1}^\prime\right)\times \Gamma_{Z_2}\left(\lambda_{Z_2},\lambda_{Z_2}^\prime\right),
		\end{aligned}
	\end{equation}
	where $\Gamma^\prime$s are the normalized decay density matrix given in the Appendix~\ref{sec:sdm}. For the single $Z$ boson production case, the angular distribution of final state leptons is given by~\cite{Boudjema:2009fz},
	\begin{equation}
		\begin{aligned}
			\frac{1}{\sigma}\frac{d\sigma}{d\Omega} &= \frac{3}{4\pi} \Bigg[ \frac{1}{6} + \frac{\delta}{2} + \frac{1}{2}\alpha \sin\theta (a_1 \cos\phi + a_2 \sin\phi)
			 + \frac{1}{2}\alpha a_3 \cos\theta + \frac{(1 - 3\delta)}{4} a_4 \sin^2\theta \sin(2\phi) \\
			& + \frac{(1 - 3\delta)}{4} \sin(2\theta) (a_5 \cos\phi + a_6 \sin\phi) 
			 + \frac{(1 - 3\delta)}{4} a_7 \sin^2\theta \cos(2\phi) 
			 \\&+ \frac{(1 - 3\delta)}{8\sqrt{3}} a_8 (1 + 3 \cos(2\theta)) \Bigg].
		\end{aligned}
	\end{equation}
	Here, $\theta$ and $\phi$ are the polar and azimuthal angles of the final decayed leptons in the rest frame of the $Z$ boson with their would-be momentum along $z$-direction. One can construct several asymmetries related to the polarization parameters, $a_i$, by partially integrating the angular distributions with respect to $\theta$ and $\phi$. For example, the asymmetries related to the vector polarization are given by~\cite{Rahaman:2021fcz}
	\begin{equation}
		\label{eq:polasymm}
		\begin{aligned}
			\mathcal{A}_1 &= \left(\int_{\theta=0}^\pi\int_{\phi = 0}^{\pi/2}-\int_{\theta=0}^\pi\int_{\phi=\pi/2}^{3\pi/2}+\int_{\theta=0}^\pi\int_{\phi=3\pi/2}^{2\pi}\right)d\Omega \left(\frac{1}{\sigma}\frac{d\sigma}{d\Omega}\right)\\
			&\equiv \frac{\sigma(\sin\theta\cos\phi >0)-\sigma(\sin\theta\cos\phi < 0)}{\sigma(\sin\theta\cos\phi >0)+\sigma(\sin\theta\cos\phi < 0)}\\
			&= \frac{3}{4}\alpha a_1,\\
			\mathcal{A}_2 &= \left(\int_{\theta=0}^\pi\int_{\phi=0}^{\pi}-\int_{\theta=0}^\pi\int_{\phi=\pi}^{2\pi}\right)d\Omega\left(\frac{1}{\sigma}\frac{d\sigma}{d\Omega}\right)\\
			&\equiv \frac{\sigma(\sin\theta\sin\phi >0)-\sigma(\sin\theta\sin\phi < 0)}{\sigma(\sin\theta\sin\phi >0)+\sigma(\sin\theta\sin\phi < 0)}\\
			&= \frac{3}{4}\alpha a_2,\\
			\mathcal{A}_3 &= \left(\int_{\theta=0}^{\pi/2}\int_{\phi=0}^{2\pi}-\int_{\theta=\pi/2}^\pi\int_{\phi=0}^{2\pi}\right)d\Omega\left(\frac{1}{\sigma}\frac{d\sigma}{d\Omega}\right)\\
			&\equiv \frac{\sigma(\cos\theta >0)-\sigma(\cos\theta < 0)}{\sigma(\cos\theta >0)+\sigma(\cos\theta < 0)}\\
			&= \frac{3}{4}\alpha a_3
		\end{aligned}
	\end{equation}
	The other remaining five polarization parameters can be similarly obtained through partially integrating out the differential rate. The comprehensive details can be found in Ref.~\cite{Rahaman:2021fcz}. In the case when two $Z$ bosons are co-produced, the joint correlated angular distribution of the final leptons given in Eq.~\eqref{eq:jointad} can be compactly written as
	\begin{equation}
		\frac{1}{\sigma}\frac{d^2\sigma}{d\Omega^{l_1}d\Omega^{l_2}} \propto \sum C_{ij} \times  f_i^{l_1}\cdot f_j^{l_2},~i,j \in \{0,\dots,8\}.
	\end{equation}
	Here, $C$ is a $9\times 9$ matrix where $C_{00}$ entry is non-informative, while $C_{0i}$ and $C_{j0}$ represents eight polarizations of each $Z$ boson, and $C_{ij}, i,j \ge 1$ represents $64$ spin correlations parameters of $ZZ^\star$ system. The functions $f_i^{l_1/l_2}$ are,
	\begin{equation}
		\label{eq:functions}
		\begin{aligned}
			&f_i^{l_1/l_2} = \{1, \sin\theta^{l_1/l_2}\cos\phi^{l_1/l_2},\sin\theta^{l_1/l_2}\sin\phi^{l1_/l_2},\cos\theta^{l_1/l_2},\sin^2\theta^{l_1/l_2}\sin(2\phi^{l_1/l_2}),\\&\sin(2\theta^{l_1/l_2})\cos\phi^{l_1/l_2},\sin(2\theta^{l_!/l_2})\cos\phi^{l_1/l_2},\sin(2\theta^{l_1/l_2})\sin\phi^{l_1/l_2},\\&\sin^2\theta^{l_1/l_2}\cos(2\phi^{l_1/l_2}),1+3\cos(2\theta^{l_1/l_2})\}.
		\end{aligned}
	\end{equation}
	The each elements of matrix $C$ can then be obtained as an asymmetries in functions $f^\prime$s as discussed above in Eq.~\eqref{eq:polasymm}. It would corresponds to 
    \begin{equation}
        \label{eq:connect}
        C_{ij} = M_{ij} \ A_{ij}(f),
    \end{equation}
     where $M$ is a $9\times 9$ matrix connecting asymmetries to polarizations and spin correlations. Numerically, these asymmetries can be found as a counting procedure as,
	\begin{equation}
		\mathcal{A}_{ij}(f) = \frac{\sigma(f_i^{l_1}f_j^{l_2} > 0) - \sigma(f_i^{l_1}f_j^{l_2})}{\sigma(f_i^{l_1}f_j^{l_2} > 0) + \sigma(f_i^{l_1}f_j^{l_2})},
	\end{equation}
	where $f_i^{l_1/l_2}$ are eight angular functions associated with final decayed leptons from two different $Z$ bosons listed in Eq.~\eqref{eq:functions}. For $i=0$, the $\mathcal{A}$ depicts eight polarizations asymmetries of $Z_2$ boson, while for $j=0$ the $\mathcal{A}$ gives polarizations asymmetries of $Z_1$ boson. And the rest $\mathcal{A}_{ij}, i,j \ge 1$ relates to asymmetries associated with $64$ spin-correlations. Once, these asymmetries are found, converting these into parameters of density matrix remains straightforward provided a proportionality factor. We will discuss the quantum tomography of $ZZ^\star$ system in the next section.
\section{Entanglement in $ZZ^\star$ system}
	\label{sec:ent}
    	Quantum entanglement plays a key role in quantum information resources, see~\cite{Horodecki:2009ZZ,Yu_2021}. Thus, characterization and quantification of entanglement have become an important problem in quantum information science. There are several entanglement measure (EM) proposed in the current literature for bipartite system such as the entanglement of formation (EOF)~\cite{bennett1996mixed}, concurrence~\cite{wootters1998entanglement}, relative entropy~\cite{vedral1997quantifying}, geometric entanglement~\cite{wei2003geometric}, negativity~\cite{vidal2002computable} and squashed entanglement~\cite{christandl2004squashed,yang2009squashed}. Among them, EOF is one of the most famous measures of entanglement. For a pure bipartite state $\ket{\psi}_{AB}$ in the Hilbert space, the EOF is given by
	\begin{equation}
		E_F\left(\ket{\psi}_{AB}\right) = S(\rho_A),
	\end{equation}
	where $S(\rho_A) = - \text{Tr}\left[\rho_A \text{log}_2\rho_A\right]$ is the von Neumann entropy of the reduced density matrix of the system $A$. For a bipartite mixed states $\rho_{AB}$, the EOF is defined by the convex roof
	\begin{equation}
		E_F\left(\rho_{AB}\right) = \text{min}\sum_i p_i E_F\left(\ket{\psi_i}_{AB}\right),
	\end{equation}
	where the minimum is taken over all possible pure state decomposition of $\rho_{AB} = \sum_i p_i \ket{\psi_i}_{AB}\bra{\psi_i}$ with $\sum_ip_i = 1$ and $p_i > 0$. The EOF provides an upper bound on the rate at which maximally entangled states can be distilled from $\rho$ and a lower bound on the rate at which maximally entangled states are needed to prepare copies of $\rho$~\cite{hayden2001asymptotic}. Ref.~\cite{wootters1998entanglement} provided a formula for EOF in the case of a two-qubit system as,
	\begin{equation}
		E_F(\rho) = \mathcal{E}\left(C\left(\rho\right)\right),
	\end{equation}
	where the concurrence is given by $C(\rho) = \text{max}\{0,\lambda_1-\lambda_2-\lambda_3-\lambda_4\}$ with $\lambda_i^\prime$s are the eigenvalues, in decreasing order, of the Hermitian matrix $R \equiv \sqrt{\sqrt{\rho}\widetilde{\rho}\sqrt{\rho}}$. The spin-flipped state is defined as $\widetilde{\rho} = (\sigma_y \otimes \sigma _y)\rho^\star (\sigma_y \otimes \sigma_y)$, where $\sigma_y$ is the second Pauli matrix. The function $\mathcal{E}$ is given by~\cite{wootters1998entanglement} as,
	\begin{equation}
		\begin{aligned}
			\mathcal{E}(C) &= h\left(\frac{1+\sqrt{1-C^2}}{2}\right),\\
			h(x) &= -x\text{log}_2 x - (1-x)\text{log}_2(1-x).
		\end{aligned}
	\end{equation}
    In our current work, we employ the concurrence measure to highlight the entanglement in the $ZZ^\star$ system. Such measure can be translated to the more general form of EOF such as Renyi-$\alpha$ entropy for $3\otimes 3$ quantum states as shown in Ref.~\cite{Song_2016}. For a pure bipartite state, the concurrence is given by $C\left(\ket{\psi}\right) = \sqrt{2\left(1-\text{Tr}\rho_A^2\right)}$, while for the mixed bipartite state, it is defined by convex roof $C(\rho) = \sum_i p_i C\left(\ket{\psi_i}\right)$ for all possible pure state decomposition of $\rho = \sum_i p_i \ket{\psi_i}\bra{\psi_i}$. Refs.~\cite{chen2005concurrence,mintert2007observable,zhang2008observable, Zhu_2012} has obtained a series of lower and upper bounds for concurrence. The upper bound on concurrence is given by~\cite{Zhu_2012}
	\begin{equation}
		\label{eq:ub}
		\mathcal{C}[\rho]_{\text{UB}} = \text{min}\left[\sqrt{2[1-\text{Tr}\rho_a^2]},\sqrt{2[1-\text{Tr}\rho_2^2]}\right].
	\end{equation} 
	The lower bound for concurrence for a bipartite system of arbitrary dimension is given by~\cite{Song_2016}
	\begin{equation}
		\label{eq:lb}
		\begin{aligned}
			&\mathcal{C}^2[\rho]_{\text{LB}} = \text{max}\left[0,~\frac{2}{m(m-1)}(|\rho^T|_1-1)^2,\frac{2}{m(m-1)}(|R(\rho)|_1-1)^2,\right.\\&\left.2[\text{Tr}\rho^2-\text{Tr}\rho_1^2],~2[\text{Tr}\rho^2-\text{Tr}\rho_2^2]\right],
		\end{aligned}
	\end{equation}
	with $\rho^T$ is partially transposed density matrix and $R(\rho)$ is realigned density matrix, such that, $R(\rho)_{ij,kl} = \rho_{ik,jl}$ where $i$ and $j$ are row and column of sub-system $\rho_1$, while $k$ and $l$ are row and column indices for sub-system $\rho_2$. The $\rho_{1/2}$ are the reduced density matrix representing two $Z$ bosons. The sub-system $\rho_{1/2}$ are obtained by doing a partial trace of complete $\rho$ as $\rho_{1/2} = \text{Tr}_{2/1}\rho$. The $|.|_1$ denotes the trace norm defined as $|\rho| = \text{Tr}\left[\sqrt{\rho^\dagger \rho}\right]$. The non-zero positive value of lower bounds indicates that the states exist in an entangled state. 
    	
	At this juncture, we would like to highlight the importance of the symmetry factor that arises because of fundamental limits on identifying two $Z$ bosons to be distinct objects even with different final flavor objects. It has been earlier reported in~\cite{Ashby-Pickering:2022umy} that the use of un-symmetrized density matrix leads to negative probabilities and eigenvalues, making the whole business of finding entanglement measures null and void. In Ref.~\cite{Ashby-Pickering:2022umy}, the symmetrized parameters of the density matrix is given in the form of,
	\begin{equation}
		\begin{aligned}
			\hat{a}_i &= \hat{b}_i = \frac{1}{8}\left\langle \widetilde{\Phi}^P_{F_l^{(1)},i}(\hat{n}_1) + \widetilde{\Phi}^P_{F_l^{(2)},i}(\hat{n}_2)\right\rangle_{\mathrm{av}},\\
			\hat{c}_{ij} &= \hat{c}_{ji} = \frac{1}{16}\left\langle \widetilde{\Phi}^P_{F_l^{(1)},i}(\hat{n}_1)\widetilde{\Phi}^P_{F_m^{(2)},j}(\hat{n}_2) + i \longleftrightarrow j\right\rangle,
		\end{aligned}
	\end{equation}
	where $\widetilde{\Phi}^P$ are the angular functions associated with final leptons, see Ref.~\cite{Ashby-Pickering:2022umy} for details. In our case, the above discussion would translate for finding the asymmetries and the proportionality constant as follows; for eight polarizations the angular functions would be $f_i^{e^-}+f_i^{\mu^-}$, while for $64$ spin-correlations, the symmetric functions would be $f_{i}^{e^-}f_j^{\mu^-}+f_j^{e^-}f_i^{\mu^-}$ as opposed to $f_i^{e^-}f_j^{\mu^-}$. In a compact way, we find the parameters as
	\begin{equation}
		\label{eq:symmjdm}
		\begin{aligned}
			a_i = b_i \propto \frac{1}{\sigma}\int d\Omega \frac{d\sigma}{d\Omega}\mathrm{Sign}\left(f_i+f_j\right).
		\end{aligned}
	\end{equation}
	In this approach, we do not have to worry about the additional factor of $1/2$ which have been introduced as ad-hoc parameter in~\cite{Ashby-Pickering:2022umy} to make a valid density matrix. The only thing that remain is the calculation of proportionality factor or matrix $M$ that connects asymmetries to polarizations and spin correlations parameters, as given in Eq.~\eqref{eq:connect}. For that, we perform a numerical integration of symmetrized version of Eq.~\eqref{eq:jointad} as shown above in Eq.~\eqref{eq:symmjdm}. We list these factors in Appendix~\ref{sec:polfact}. We have checked that the problem of negative probabilities and eigenvalues disappears using symmetric version, and is depicted in Appendix~\ref{sec:negdm}.
    \begin{figure*}[!h]
		\centering
		\includegraphics[width=0.49\textwidth]{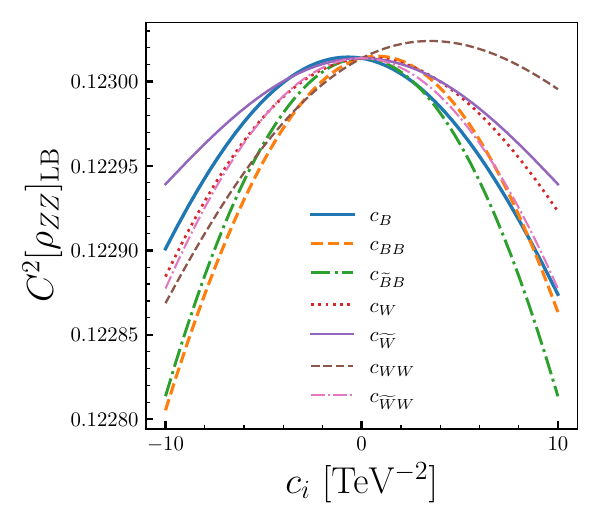}
		\includegraphics[width=0.49\textwidth]{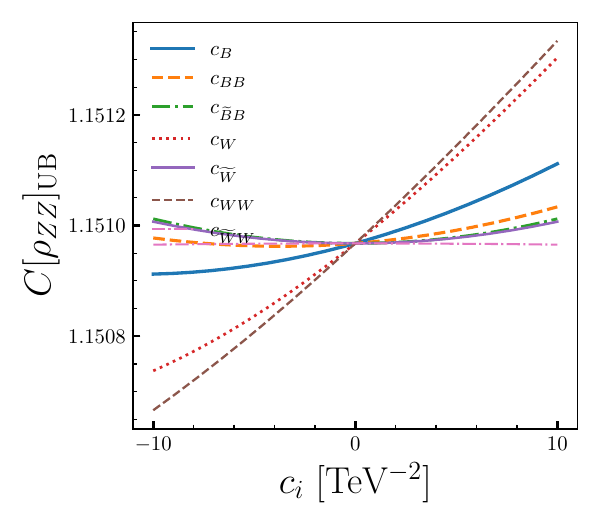}
		\caption{Variation of lower bound $C^2[\rho_{ZZ}]_{\mathrm{LB}}$ (left panel) and upper bound $C[\rho_{ZZ}]_{\mathrm{UB}}$ for the concurrence as a function of anomalous couplings in $ZZ^\star$ system. One couplings are varied at a time while others are explicitly kept to zero.}
		\label{fig:lbanom}
	\end{figure*}
    
	In Fig.~\ref{fig:lbanom}, we show the distribution of lower (left panel) and upper (right panel) bounds as a function of one anomalous coupling at a time for $ZZ^\star$ state produced from the Higgs boson decay. We note that for all values of anomalous couplings $c_i$, the $ZZ^\star$ states are entangled. In the case of $CP$-odd couplings, the maximal value of the lower bound for concurrence reaches $0.123$ for $c_i = 0.0$, which is the SM point. Meanwhile, for the $CP$-even couplings, the maximum lower bound is shifted from the SM point. The maximum exist for $-0.7,0.8,0.4,$ and $3.5$, respectively for $c_B$, $c_{BB}$, $c_W$, and $c_{WW}$ couplings. These highlight the sensitivity of the lower bound on the search for deviation from the SM prediction. The upper bound on the concurrence shows a variation with anomalous couplings. For a pure state, the upper bound following in Eq.~\eqref{eq:ub} is zero, and for the maximally mixed state, it reaches $1.4$. Though the upper bound does not directly guarantee the states will be entangled, it could point to the degree of $ZZ^\star$ being in a mixed state. Moreover, the distribution points that the $ZZ^\star$ states exist to be in a mixed state.
    	\begin{figure*}[!h]
		\centering
		\includegraphics[width=0.49\linewidth]{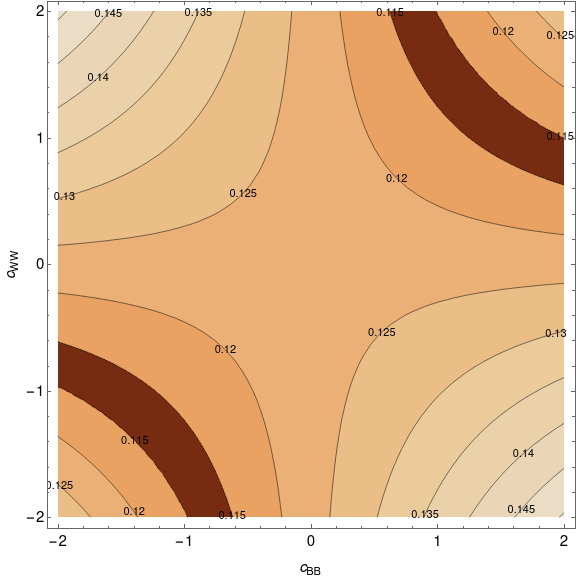}
		\includegraphics[width=0.49\linewidth]{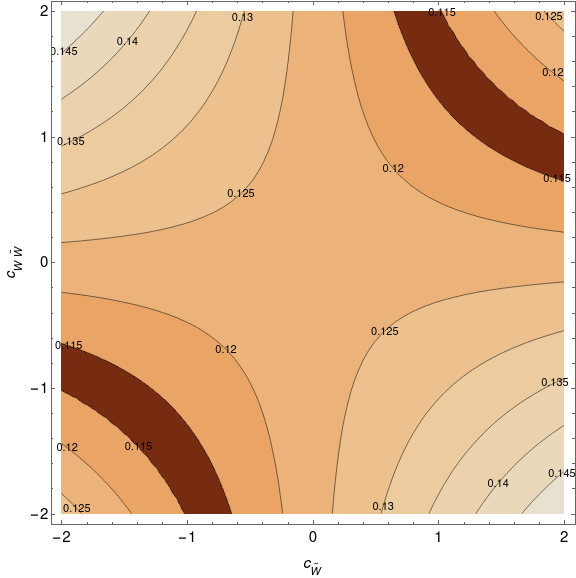}
		\caption{Two dimensional contour plots highlighting the lower bound $C^2[\rho_{ZZ}]_{\mathrm{LB}}$ for concurrence as a function of two anomalous couplings. In the left panel, we show the distribution for both CP-even couplings and in the right panel, we show distribution of lower bound for both CP-odd couplings.}
		\label{fig:twod1}
	\end{figure*}	
    
\begin{figure*}[!h]
		\centering
		\includegraphics[width=0.49\linewidth]{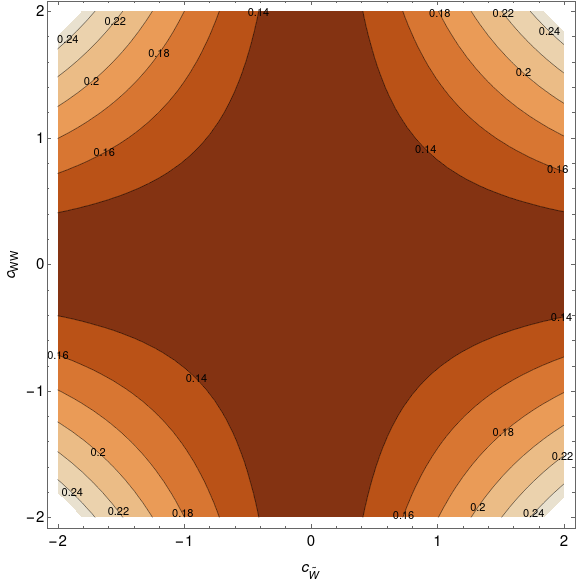}
		\includegraphics[width=0.49\linewidth]{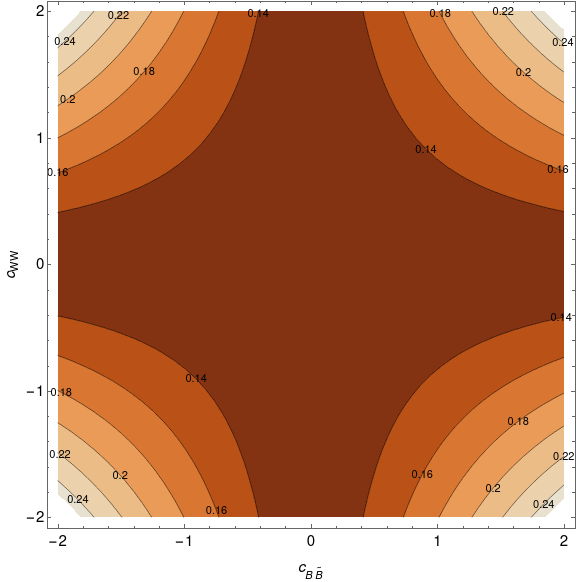}
		\caption{Two dimensional contour plots highlighting the lower bound $C^2[\rho_{ZZ}]_{\mathrm{LB}}$ for concurrence as a function of CP-mixed anomalous couplings. In the right panel we show the distribution for $c_{\widetilde{W}}-c_{WW}$ and in the left panel depicts lower bound for $c_{\widetilde{B}B} - c_{WW}$ couplings.}
		\label{fig:twod2}
	\end{figure*}

\begin{figure*}[!h]
		\centering
		\includegraphics[width=0.49\textwidth]{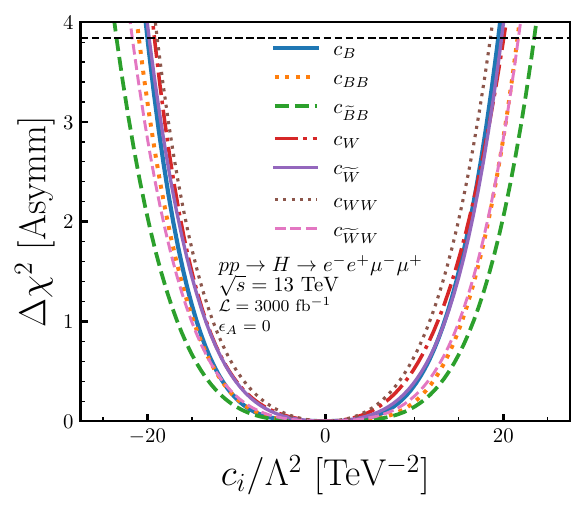}
		\includegraphics[width=0.49\textwidth]{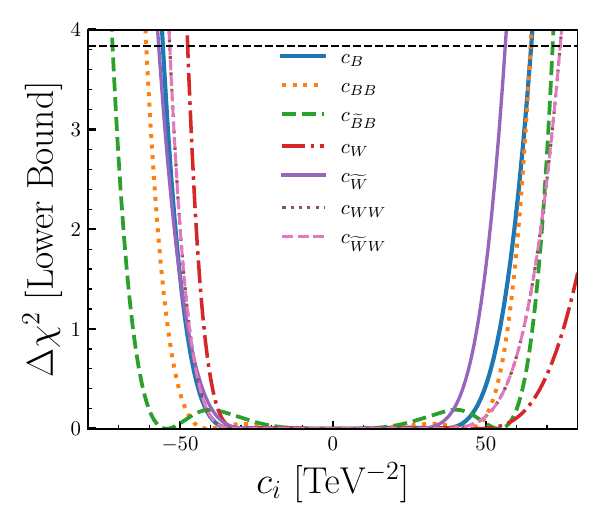}
		\caption{\label{fig:chi2lb}Chi-squared dependence for asymmetries and lower bound of concurrence as a function of one anomalous coupling at a time. The distribution are obtained for $pp\to H \to e^-e^+\mu^-\mu^+$ process at $\sqrt{s}=13$ TeV and integrated luminosity of $3000$ fb$^{-1}$. The systematic errors are not considered for this analysis.}
	\end{figure*}

	Next, we examine the variation in lower bound (\( C^2[\rho_{ZZ^\star}]_{\mathrm{LB}} \)) in scenarios where two anomalous couplings are non-zero simultaneously. The results are presented in Figs.~\ref{fig:twod1} and \ref{fig:twod2}, where we illustrate two-dimensional contour plots of the lower bounds on concurrence as a function of these paired anomalous couplings. For clarity, all other couplings are set explicitly to zero to isolate the effects of the two couplings in focus. In the instances where both anomalous couplings are CP-even or CP-odd (see Fig.~\ref{fig:twod1}), we observe that the lower bound values start at an intermediate level and progressively increases as the magnitudes of the anomalous couplings rise in the second and fourth quadrants. However, along the \( x = y \) axis, the lower bounds tend to decrease until the anomalous couplings reach relatively large values. This behavior results from cancellations in interference terms between the two anomalous couplings and the SM amplitudes, thereby reducing the $C^2[\rho_{ZZ^\star}]_{\mathrm{LB}}$ in this region. 
In contrast, when CP-mixed couplings are considered (as depicted in Fig.~\ref{fig:twod2}), such as the pairs \( (c_{\widetilde{B}B}, c_{WW}) \) and \( (c_{\widetilde{W}}, c_{WW}) \), we find minimal values of the \( C^2[\rho_{ZZ^\star}]_{\mathrm{LB}} \) when one coupling is near the SM or both anomalous couplings are near zero. Conversely, the \( C^2[\rho_{ZZ^\star}]_{\mathrm{LB}} \) rises as both anomalous couplings increase in magnitude in all quadrants. Consequently, measuring the lower bound on concurrence becomes crucial in highlighting the CP structure associated with new physics contributions.
\\\\
	And finally, we assess the sensitivity of \( C^2[\rho_{ZZ^\star}]_{\mathrm{LB}} \) to anomalous couplings. This approach allows us to rigorously constrain the coupling parameters within the framework of new physics, using the \( C^2[\rho_{ZZ^\star}]_{\mathrm{LB}} \) as a diagnostic tool. Due to the complexity of the analytic form of the lower bound, we perform numerical analysis to compute the error in the measurement of the lower bound. For each value of anomalous coupling, we change the value of asymmetries as $A_i[c]^\pm = A_i[c] \pm \delta A$, with $\delta A = \sqrt{\frac{1-A^2}{\mathcal{L}\sigma} + \epsilon_A^2}$, where $\epsilon_A$ is the systematic error in the estimation of asymmetries. The error in the lower bound is then found as
	\begin{equation}
		\Delta C^2[\rho_{ZZ^\star}]_{\mathrm{LB}} = \sqrt{\left(\frac{C^{2}[\rho_{ZZ^\star}^+]_{\mathrm{LB}}-C^2[\rho^-_{ZZ^\star}]_{\mathrm{LB}}}{2}\right)^2}. 
	\end{equation}
	We then compute the chi-squared for asymmetries as $\Delta\chi^2(c_i) = \left(\frac{\mathcal{O}(c_i)-\mathcal{O}(0)}{\delta A}\right)^2$ and we represent the result for all seven couplings with quantum observables as well as collider observables (asymmetries) in Fig.~\ref{fig:chi2lb}. From the chi-squared distribution, it becomes evident that relying solely on measuring the lower bound as an observable for probing deviations from the SM prediction leads to poorer sensitivity. 
	This is primarily due to the non-inclusion or cancellation of parameters of the density matrix, which leads to a lower bound, providing poorer information in capturing the full extent of possible deviations. In contrast, incorporating the measure of entanglement offers a more nuanced approach, allowing one to probe the quantum nature of the interacting particles. This enhances the potential for detecting subtle quantum effects, indicating new physics. However, we recommend focusing on base angular functions as observables for precision measurements aimed at setting stringent limits on new physics. These functions provide a more robust framework for extracting precise constraints on potential deviations from the SM, particularly in high-energy collider experiments.

    	\section{Conclusion}
	\label{sec:conclude}
	We study the single Higgs boson production process, $pp \to H \to e^-e^+\mu^-\mu^+$, in LHC at a center-of-mass energy of $13$ TeV in presence of anomalous $HZZ$ couplings. We consider the general $HZZ$ coupling involving both CP-even and odd parameters induced by dimension-6 operators. The density matrix corresponding to the $ZZ^\star$ system was reconstructed using the asymmetries of joint angular function of final decayed leptons. However, such density matrix suffers from negative probabilities and eigenvalues. To circumvent the negativity problem, we reconstruct the parameters of the density matrix using a symmetrized angular function owing to the indistinguishability of two $Z$ bosons.
	\\\\
	We compute the lower bound for concurrence to establish the entangled nature of the $ZZ^\star$ system. For all values of anomalous couplings, the $ZZ^\star$ system were observed to be entangled. In the case of CP-odd couplings, the maximal value of the lower bound equates to that of the SM, while for the CP-even couplings, there is a significant shift in the maximal value of the lower bound. This highlights how the lower bound for concurrence could be implemented to dissect the CP structure of new physics. For the scenario when two anomalous couplings were considered simultaneously, we noted a stark difference between the cases when both the couplings were either CP-even or odd and in CP-mixed cases. For e.g, where both anomalous couplings are CP-even or both are CP-odd, we observe that the lower bound values start at an intermediate level and progressively increase as the magnitudes of the anomalous couplings rise in the second and fourth quadrants. And in the first and fourth quadrants, the lower bounds tend to decrease until the anomalous couplings reach relatively large values. In contrast, when CP-mixed couplings are considered, we find minimal values of the lower bound when one coupling is near the SM or both couplings are close to zero. The lower bound rises as both anomalous couplings increase in magnitude in all four quadrants.
	\\\\
	The chi-squared analysis for lower bound shows a poorer sensitivity compared to that of asymmetries which were used to reconstruct the density matrix. Measuring the lower bound becomes essential to highlight the quantum nature of interacting particles. However, for performing a precision study of new physics, the use of parameters of density matrix (i.e. asymmetries) directly becomes inevitable. 
\appendix
\section{Normalized decay density matrix}
	\label{sec:sdm}
	\noindent
	For the decay of spin-1 $Z$ boson decaying to $f\bar{f}$ pairs with decay vertex $\bar{f}\gamma^\mu P_L f V$, the decay density matrix is given by~\cite{Boudjema:2009fz},
		\begin{equation}
			\Gamma_Z(\lambda_Z,\lambda_Z^\prime) =
			\begin{bmatrix}
				\frac{1+\delta+(1-3\delta)\cos^2\theta+2\alpha\cos\theta}{4}&\frac{\sin\theta(\alpha+(1-3\delta)\cos\theta)}{2\sqrt{2}}e^{i\phi}&(1-3\delta)\frac{(1-\cos^2\theta)}{4}e^{i2\phi}\\\\
				\frac{\sin\theta(\alpha+(1-3\delta)\cos\theta)}{2\sqrt{2}}e^{-i\phi}&\delta+(1-3\delta)\frac{\sin^2\theta}{2}&\frac{\sin\theta(\alpha-(1-3\delta)\cos\theta)}{2\sqrt{s}}e^{i\phi}\\\\
				(1-3\delta)\frac{(1-\cos^2\theta)}{4}e^{-i2\phi}&\frac{\sin\theta(\alpha-(1-3\delta)\cos\theta)}{2\sqrt{2}}e^{-i\phi}&\frac{1+\delta+(1-3\delta)\cos^2\theta-2\alpha\cos\theta}{4}
			\end{bmatrix},
		\end{equation}
	where the $\theta$, and $\phi$ are the polar and azimuth orientation of final decayed fermions at the rest frame of $Z$ boson. Here the spin analyzing power is given by~\cite{Boudjema:2009fz}
		\begin{equation*}
			\alpha = \frac{2(C_R^2-C_L^2)\sqrt{1+(x_1^2-x_2^2)^2-2(x_1^2+x_2^2)}}{12C_LC_Rx_1x_2+(C_R^2+C_L^2)\left[2-(x_1^2-x_2^2)^2+(x_1^2+x_2^2)\right]}.
	\end{equation*}
	And the parameter \(\delta\) for the case \(Z\) boson decaying to two fermions is defined as~\cite{Boudjema:2009fz}
		\begin{equation*}
			\delta = \frac{4C_LC_Rx_1x_2+(C_R^2+C_L^2)\left[(x_1^2+x_2^2)-(x_1^2-x_2^2)^2\right]}{12C_LC_Rx_1x_2+(C_R^2+C_L^2)\left[2-(x_1^2-x_2^2)^2+(x_1^2+x_2^2)\right]},
	\end{equation*}
	where \(x_i = m_i/M\) with \(m_i\) as the mass of the final jets and \(M\) as the mass of the \(Z\) boson. At the high-energy limit, \(x_i \to 0\), and \(\alpha \to (C_R^2-C_L^2)/(C_R^2+C_L^2)\), and \(\delta \to 0\). Within the SM at leading order, we found for the \(Z\to l^-l^+\) decay, \(\alpha \approx 0.216\).

\section{The spin-1 operator and $J$ basis}
	\label{sec:jbasis}
	The three spin-1 operators are $3\times 3$ matrix defined as,
	\begin{equation}
		\begin{aligned}
			S_x = \frac{1}{\sqrt{2}}\begin{pmatrix}
				0&1&0\\1&0&1\\0&1&0
			\end{pmatrix},~S_y = \frac{1}{\sqrt{2}}\begin{pmatrix}
				0&-i&0\\i&0&-i\\0&i&0
			\end{pmatrix}
			S_z = \begin{pmatrix}
				1&0&0\\0&0&0\\0&0&-1
			\end{pmatrix}.
		\end{aligned}
	\end{equation}
	We construct the $J_i$ basis used to define the spin-1 density matrix as,
	\begin{equation}
		\begin{aligned}
			J_i &= S_i/2,~J_4 = (S_x.S_y + S_y.S_x)/2\\
			J_5 &= (S_x.S_z + S_z.S_x)/2,~J_6 = (S_y.S_z + S_z.S_y)/2\\
			J_7 &= (S_x.S_x - S_y.S_y)/2,~J_8 = \sqrt{3}(S_z.S_z/2 - \mathbf{I}/3)        
		\end{aligned}
	\end{equation}
	The $J_i$ matrices satisfy follows normalization condition, $\text{Tr}[J_i,J_j] = \delta_{ij}/2$.
\section{Relation between symmetrized asymmetries and parameters of density matrix}
	\label{sec:polfact}
	\noindent
	Since, the two $Z$ bosons are fundamentally identical, it becomes essential to consider the symmetrized functions while constructing the joint density matrix form the angular information of final decayed fermions. Due to complexity in the analytic form of symmetrized angular function, we used numerical integration technique to obtain the proportionality factor connecting asymmetries with density matrix parameters. In general, each polarizations and spin-correlations are related to asymmetries can be denoted in a matrix form as,
	$ C = M^{-1} A$. Here, $C,M,$ and $A$ are $9\times 9$ matrix. The matrix $A$ contains $80$ asymmetries where the first element is just a identity and remaining 8 elements of first row and columns denotes asymmetries associated with polarizations of two $Z$ bosons. The remaining 64 elements corresponds to spin-correlation asymmetries. The matrix $M$ encodes the proportionality factor connecting asymmetries and spin matrices.
		\begin{equation}
			M = \begin{bmatrix}
				1.000&0.216&0.216&0.216&0.444&0.445&0.445&0.445&0.453\\\\
				0.216&0.053&0.041&0.065&0.070&0.070&0.085&0.076&0.082\\\\
				0.216&0.041&0.053&0.060&0.070&0.085&0.070&0.076&0.086\\\\
				0.216&0.065&0.060&0.000&0.052&0.052&0.052&0.085&0.079\\\\
				0.444&0.070&0.070&0.052&0.203&0.159&0.159&0.159&0.166\\\\
				0.445&0.070&0.085&0.052&0.159&0.203&0.159&0.166&0.169\\\\
				0.445&0.085&0.070&0.052&0.159&0.159&0.203&0.166&0.170\\\\
				0.445&0.076&0.076&0.085&0.159&0.166&0.166&0.203&0.166\\\\
				0.453&0.082&0.086&0.079&0.166&0.169&0.170&0.166&0.222
			\end{bmatrix}
		\end{equation}
\section{Negativity in the symmetric density matrix}
	\label{sec:negdm}
	The reconstructed density matrix for the non-symmetric $\rho_{ZZ^\star}^{\mathrm{Non-symm}}$ is obtained as
 \begin{center}
\begin{sideways}
$\left(
\begin{array}{ccccccccc}
0.2 & 0.012+0.004 i & -0.003+0.003 i & 0.005+0.008 i & -0.314-0.006 i & 0.001-0.012 i & 0.003+0.003 i & 0.001 & 0.201-0.003 i \\\\
\text{} & 0.143 & 0.006-0.001 i & -0.012-0.008 i & 0.008 i & -0.011+0.006 i & -0.007-0.007 i & 0.003-0.001 i & -0.007+0.003 i \\\\
\text{} & \text{} & -0.01 & -0.006-0.015 i & -0.008-0.004 i & -0.005-0.014 i & 0.001+0.007 i & -0.006-0.009 i & -0.008+0.002 i \\\\
\text{} & \text{} & \text{} & 0.131 & 0.003+0.007 i & -0.002+0.003 i & -0.002+0.003 i & -0.015-0.001 i & 0.003-0.011 i \\\\
\text{} & \text{} & \text{} & \text{} & 0.121 & 0.001+0.007 i & -0.017-0.007 i & -0.001+0.009 i & -0.330+0.008 i \\\\
\text{} & \text{} & \text{} & \text{} & \text{} & 0.080 & -0.003-0.012 i & -0.009-0.01i & 0.005-0.009 i \\\\
\text{} & \text{} & \text{} & \text{} & \text{} & \text{} & -0.001 & -0.013-0.01i & 0.005-0.004 i \\\\
\text{} & \text{} & \text{} & \text{} & \text{} & \text{} & \text{} & 0.069 & -0.008-0.007 i \\\\
\text{} & \text{} & \text{} & \text{} & \text{} & \text{} & \text{} & \text{} & 0.266 \\\\
\end{array}
\right)$
\end{sideways}
\end{center}

	The eigenvalues of the non-symmetric density matrix are,
	$$0.762, -0.206, 0.158, 0.126, 0.091, 0.072, 0.023, -0.018, -0.007.$$ And the symmetric density matrix is found to be,
	\begin{equation}
		\centering
		\rho_{ZZ^\star}^{\text{Symm}}=
		    \left(
		\begin{array}{ccccccccc}
			0.095 & 0.0 & 0.0 & 0.0 & -0.003 & 0.0 & 0.0 & 0.0 & 0.004 \\\\
			\text{} & 0.116 & 0.0 & 0.0 & 0.0 & 0.003 & 0.0 & 0.0 & 0.0\\\\
			\text{} & \text{} & 0.095 & 0.0& 0.0 & 0.0& 0.0 & 0.0 & 0.0\\\\
			\text{} & \text{} & \text{} & 0.116 & 0.0 & 0.0 & 0.0& 0.003 & 0.0\\ \\
			\text{} & \text{} & \text{} & \text{} & 0.154 & 0.0 & 0.0 & 0.0 & -0.003\\ \\
			\text{} & \text{} & \text{} & \text{} & \text{} & 0.116 &0.0 & 0.0 & 0.0 \\\\
			\text{} & \text{} & \text{} & \text{} & \text{} & \text{} & 0.095 & 0.0 & 0.0 \\\\
			\text{} & \text{} & \text{} & \text{} & \text{} & \text{} & \text{} & 0.116 & 0.0 \\\\
			\text{} & \text{} & \text{} & \text{} & \text{} & \text{} & \text{} & \text{} & 0.095 \\\\
		\end{array}
		\right)
	\end{equation}
	The eigenvalues of the symmetric density matrix given above are,
	$$0.155, 0.119, 0.119, 0.1134, 0.113, 0.099,0.095, 0.095, 0.091$$.
\acknowledgments

 A. Subba acknowledges the financial support of the University Grants Commission, Government of India, through the UGC-NET Senior Research Fellowship.


\bibliographystyle{JHEP}
\bibliography{biblio.bib}

\providecommand{\href}[2]{#2}\begingroup\raggedright\begin{thebibliography}{10}

\bibitem{PhysRev.47.777}
A.~Einstein, B.~Podolsky and N.~Rosen, \emph{Can quantum-mechanical description
  of physical reality be considered complete?},
  \href{https://doi.org/10.1103/PhysRev.47.777}{\emph{Phys. Rev.} {\bfseries
  47} (1935) 777}.

\bibitem{Shor:1995hbe}
P.W.~Shor, \emph{{Scheme for reducing decoherence in quantum computer memory}},
  \href{https://doi.org/10.1103/physreva.52.r2493}{\emph{Phys. Rev. A}
  {\bfseries 52} (1995) R2493}.

\bibitem{Steane:1996ghp}
A.M.~Steane, \emph{{Error Correcting Codes in Quantum Theory}},
  \href{https://doi.org/10.1103/physrevlett.77.793}{\emph{Phys. Rev. Lett.}
  {\bfseries 77} (1996) 793}.

\bibitem{Bennett:1992tv}
C.H.~Bennett, G.~Brassard, C.~Crepeau, R.~Jozsa, A.~Peres and W.K.~Wootters,
  \emph{{Teleporting an unknown quantum state via dual classical and
  Einstein-Podolsky-Rosen channels}},
  \href{https://doi.org/10.1103/PhysRevLett.70.1895}{\emph{Phys. Rev. Lett.}
  {\bfseries 70} (1993) 1895}.

\bibitem{Bennett:1992zzb}
C.H.~Bennett and S.J.~Wiesner, \emph{{Communication via one- and two-particle
  operators on Einstein-Podolsky-Rosen states}},
  \href{https://doi.org/10.1103/PhysRevLett.69.2881}{\emph{Phys. Rev. Lett.}
  {\bfseries 69} (1992) 2881}.

\bibitem{Ekert:1991zz}
A.K.~Ekert, \emph{{Quantum cryptography based on Bell's theorem}},
  \href{https://doi.org/10.1103/PhysRevLett.67.661}{\emph{Phys. Rev. Lett.}
  {\bfseries 67} (1991) 661}.

\bibitem{Jennewein:2000zz}
T.~Jennewein, C.~Simon, G.~Weihs, H.~WeinfurterD and A.~Zeilinger,
  \emph{{Quantum Cryptography with Entangled Photons}},
  \href{https://doi.org/10.1103/PhysRevLett.84.4729}{\emph{Phys. Rev. Lett.}
  {\bfseries 84} (2000) 4729}
  [\href{https://arxiv.org/abs/quant-ph/9912117}{{\ttfamily
  quant-ph/9912117}}].

\bibitem{PhysRevLett.49.91}
A.~Aspect, P.~Grangier and G.~Roger, \emph{Experimental realization of
  einstein-podolsky-rosen-bohm gedankenexperiment: A new violation of bell's
  inequalities}, \href{https://doi.org/10.1103/PhysRevLett.49.91}{\emph{Phys.
  Rev. Lett.} {\bfseries 49} (1982) 91}.

\bibitem{Hensen:2015ccp}
B.~Hensen et~al., \emph{{Loophole-free Bell inequality violation using electron
  spins separated by 1.3 kilometres}},
  \href{https://doi.org/10.1038/nature15759}{\emph{Nature} {\bfseries 526}
  (2015) 682} [\href{https://arxiv.org/abs/1508.05949}{{\ttfamily
  1508.05949}}].

\bibitem{PhysRevLett.115.250401}
M.~Giustina, M.A.M.~Versteegh, S.~Wengerowsky, J.~Handsteiner, A.~Hochrainer,
  K.~Phelan et~al., \emph{Significant-loophole-free test of bell's theorem with
  entangled photons},
  \href{https://doi.org/10.1103/PhysRevLett.115.250401}{\emph{Phys. Rev. Lett.}
  {\bfseries 115} (2015) 250401}.

\bibitem{Barr:2022wyq}
A.J.~Barr, P.~Caban and J.~Rembieli\'nski, \emph{{Bell-type inequalities for
  systems of relativistic vector bosons}},
  \href{https://doi.org/10.22331/q-2023-07-27-1070}{\emph{Quantum} {\bfseries
  7} (2023) 1070} [\href{https://arxiv.org/abs/2204.11063}{{\ttfamily
  2204.11063}}].

\bibitem{Aguilar-Saavedra:2022mpg}
J.A.~Aguilar-Saavedra, \emph{{Laboratory-frame tests of quantum entanglement in
  H\textrightarrow{}WW}},
  \href{https://doi.org/10.1103/PhysRevD.107.076016}{\emph{Phys. Rev. D}
  {\bfseries 107} (2023) 076016}
  [\href{https://arxiv.org/abs/2209.14033}{{\ttfamily 2209.14033}}].

\bibitem{Altakach:2022ywa}
M.M.~Altakach, P.~Lamba, F.~Maltoni, K.~Mawatari and K.~Sakurai, \emph{{Quantum
  information and CP measurement in
  H\textrightarrow{}\ensuremath{\tau}+\ensuremath{\tau}- at future lepton
  colliders}}, \href{https://doi.org/10.1103/PhysRevD.107.093002}{\emph{Phys.
  Rev. D} {\bfseries 107} (2023) 093002}
  [\href{https://arxiv.org/abs/2211.10513}{{\ttfamily 2211.10513}}].

\bibitem{Cheng:2023qmz}
K.~Cheng, T.~Han and M.~Low, \emph{{Optimizing fictitious states for Bell
  inequality violation in bipartite qubit systems with applications to the
  tt\textasciimacron{} system}},
  \href{https://doi.org/10.1103/PhysRevD.109.116005}{\emph{Phys. Rev. D}
  {\bfseries 109} (2024) 116005}
  [\href{https://arxiv.org/abs/2311.09166}{{\ttfamily 2311.09166}}].

\bibitem{Han:2023fci}
T.~Han, M.~Low and T.A.~Wu, \emph{{Quantum entanglement and Bell inequality
  violation in semi-leptonic top decays}},
  \href{https://doi.org/10.1007/JHEP07(2024)192}{\emph{JHEP} {\bfseries 07}
  (2024) 192} [\href{https://arxiv.org/abs/2310.17696}{{\ttfamily
  2310.17696}}].

\bibitem{Dong:2023xiw}
Z.~Dong, D.~Gon\c{c}alves, K.~Kong and A.~Navarro, \emph{{Entanglement and Bell
  inequalities with boosted tt\textasciimacron{}}},
  \href{https://doi.org/10.1103/PhysRevD.109.115023}{\emph{Phys. Rev. D}
  {\bfseries 109} (2024) 115023}
  [\href{https://arxiv.org/abs/2305.07075}{{\ttfamily 2305.07075}}].

\bibitem{Varma:2023gwh}
M.~Varma and O.K.~Baker, \emph{{Quantum entanglement in top quark pair
  production}},
  \href{https://doi.org/10.1016/j.nuclphysa.2023.122795}{\emph{Nucl. Phys. A}
  {\bfseries 1042} (2024) 122795}
  [\href{https://arxiv.org/abs/2306.07788}{{\ttfamily 2306.07788}}].

\bibitem{Barr:2024djo}
A.J.~Barr, M.~Fabbrichesi, R.~Floreanini, E.~Gabrielli and L.~Marzola,
  \emph{{Quantum entanglement and Bell inequality violation at colliders}},
  \href{https://doi.org/10.1016/j.ppnp.2024.104134}{\emph{Prog. Part. Nucl.
  Phys.} {\bfseries 139} (2024) 104134}
  [\href{https://arxiv.org/abs/2402.07972}{{\ttfamily 2402.07972}}].

\bibitem{Fabbrichesi:2023cev}
M.~Fabbrichesi, R.~Floreanini, E.~Gabrielli and L.~Marzola, \emph{{Bell
  inequalities and quantum entanglement in weak gauge boson production at the
  LHC and future colliders}},
  \href{https://doi.org/10.1140/epjc/s10052-023-11935-8}{\emph{Eur. Phys. J. C}
  {\bfseries 83} (2023) 823}
  [\href{https://arxiv.org/abs/2302.00683}{{\ttfamily 2302.00683}}].

\bibitem{Subba:2024mnl}
A.~Subba and R.~Rahaman, \emph{{On bipartite and tripartite entanglement at
  present and future particle colliders}},
  \href{https://arxiv.org/abs/2404.03292}{{\ttfamily 2404.03292}}.

\bibitem{Morales:2023gow}
R.A.~Morales, \emph{{Exploring Bell inequalities and quantum entanglement in
  vector boson scattering}},
  \href{https://doi.org/10.1140/epjp/s13360-023-04784-7}{\emph{Eur. Phys. J.
  Plus} {\bfseries 138} (2023) 1157}
  [\href{https://arxiv.org/abs/2306.17247}{{\ttfamily 2306.17247}}].

\bibitem{Morales:2024jhj}
R.A.~Morales, \emph{{Tripartite entanglement and Bell non-locality in
  loop-induced Higgs boson decays}},
  \href{https://doi.org/10.1140/epjc/s10052-024-12921-4}{\emph{Eur. Phys. J. C}
  {\bfseries 84} (2024) 581}
  [\href{https://arxiv.org/abs/2403.18023}{{\ttfamily 2403.18023}}].

\bibitem{Fabbrichesi:2023jep}
M.~Fabbrichesi, R.~Floreanini, E.~Gabrielli and L.~Marzola, \emph{{Stringent
  bounds on HWW and HZZ anomalous couplings with quantum tomography at the
  LHC}}, \href{https://doi.org/10.1007/JHEP09(2023)195}{\emph{JHEP} {\bfseries
  09} (2023) 195} [\href{https://arxiv.org/abs/2304.02403}{{\ttfamily
  2304.02403}}].

\bibitem{Grossi:2024jae}
M.~Grossi, G.~Pelliccioli and A.~Vicini, \emph{{From angular coefficients to
  quantum observables: a phenomenological appraisal in di-boson systems}},
  \href{https://arxiv.org/abs/2409.16731}{{\ttfamily 2409.16731}}.

\bibitem{Aoude:2023hxv}
R.~Aoude, E.~Madge, F.~Maltoni and L.~Mantani, \emph{{Probing new physics
  through entanglement in diboson production}},
  \href{https://doi.org/10.1007/JHEP12(2023)017}{\emph{JHEP} {\bfseries 12}
  (2023) 017} [\href{https://arxiv.org/abs/2307.09675}{{\ttfamily
  2307.09675}}].

\bibitem{Marzola:2023oyv}
L.~Marzola, \emph{{Testing Bell inequalities and entanglement with di-boson
  final states}},  in \emph{{57th Rencontres de Moriond on Electroweak
  Interactions and Unified Theories}}, 5, 2023
  [\href{https://arxiv.org/abs/2305.08568}{{\ttfamily 2305.08568}}].

\bibitem{Abel:1992kz}
S.A.~Abel, M.~Dittmar and H.K.~Dreiner, \emph{{Testing locality at colliders
  via Bell's inequality?}},
  \href{https://doi.org/10.1016/0370-2693(92)90071-B}{\emph{Phys. Lett. B}
  {\bfseries 280} (1992) 304}.

\bibitem{ATLAS:2023fsd}
{\scshape ATLAS} collaboration, \emph{{Observation of quantum entanglement with
  top quarks at the ATLAS detector}},
  \href{https://doi.org/10.1038/s41586-024-07824-z}{\emph{Nature} {\bfseries
  633} (2024) 542} [\href{https://arxiv.org/abs/2311.07288}{{\ttfamily
  2311.07288}}].

\bibitem{CMS:2024pts}
{\scshape CMS} collaboration, \emph{{Observation of quantum entanglement in top
  quark pair production in proton\textendash{}proton collisions at $\sqrt{s} =
  13$ TeV}}, \href{https://doi.org/10.1088/1361-6633/ad7e4d}{\emph{Rept. Prog.
  Phys.} {\bfseries 87} (2024) 117801}
  [\href{https://arxiv.org/abs/2406.03976}{{\ttfamily 2406.03976}}].

\bibitem{Aguilar-Saavedra:2022uye}
J.A.~Aguilar-Saavedra and J.A.~Casas, \emph{{Improved tests of entanglement and
  Bell inequalities with LHC tops}},
  \href{https://doi.org/10.1140/epjc/s10052-022-10630-4}{\emph{Eur. Phys. J. C}
  {\bfseries 82} (2022) 666}
  [\href{https://arxiv.org/abs/2205.00542}{{\ttfamily 2205.00542}}].

\bibitem{Aoude:2022imd}
R.~Aoude, E.~Madge, F.~Maltoni and L.~Mantani, \emph{{Quantum SMEFT tomography:
  Top quark pair production at the LHC}},
  \href{https://doi.org/10.1103/PhysRevD.106.055007}{\emph{Phys. Rev. D}
  {\bfseries 106} (2022) 055007}
  [\href{https://arxiv.org/abs/2203.05619}{{\ttfamily 2203.05619}}].

\bibitem{Severi:2022qjy}
C.~Severi and E.~Vryonidou, \emph{{Quantum entanglement and top spin
  correlations in SMEFT at higher orders}},
  \href{https://doi.org/10.1007/JHEP01(2023)148}{\emph{JHEP} {\bfseries 01}
  (2023) 148} [\href{https://arxiv.org/abs/2210.09330}{{\ttfamily
  2210.09330}}].

\bibitem{Bernal:2023ruk}
A.~Bernal, P.~Caban and J.~Rembieli\'nski, \emph{{Entanglement and Bell
  inequalities violation in $H\rightarrow ZZ$ with anomalous coupling}},
  \href{https://doi.org/10.1140/epjc/s10052-023-12216-0}{\emph{Eur. Phys. J. C}
  {\bfseries 83} (2023) 1050}
  [\href{https://arxiv.org/abs/2307.13496}{{\ttfamily 2307.13496}}].

\bibitem{Bernal:2024xhm}
A.~Bernal, P.~Caban and J.~Rembieli\'nski, \emph{{Entanglement and Bell
  inequality violation in vector diboson systems produced in decays of spin-0
  particles}},  \href{https://arxiv.org/abs/2405.16525}{{\ttfamily
  2405.16525}}.

\bibitem{Sullivan:2024wzl}
M.~Sullivan, \emph{{Constraining New Physics with $h\rightarrow VV$
  Tomography}},  \href{https://arxiv.org/abs/2410.10980}{{\ttfamily
  2410.10980}}.

\bibitem{Weinberg:1978kz}
S.~Weinberg, \emph{{Phenomenological Lagrangians}},
  \href{https://doi.org/10.1016/0378-4371(79)90223-1}{\emph{Physica A}
  {\bfseries 96} (1979) 327}.

\bibitem{Weinberg:1980wa}
S.~Weinberg, \emph{{Effective Gauge Theories}},
  \href{https://doi.org/10.1016/0370-2693(80)90660-7}{\emph{Phys. Lett. B}
  {\bfseries 91} (1980) 51}.

\bibitem{Buchmuller:1985jz}
W.~Buchmuller and D.~Wyler, \emph{{Effective Lagrangian Analysis of New
  Interactions and Flavor Conservation}},
  \href{https://doi.org/10.1016/0550-3213(86)90262-2}{\emph{Nucl. Phys. B}
  {\bfseries 268} (1986) 621}.

\bibitem{Aguilar-Saavedra:2022wam}
J.A.~Aguilar-Saavedra, A.~Bernal, J.A.~Casas and J.M.~Moreno, \emph{{Testing
  entanglement and Bell inequalities in H\textrightarrow{}ZZ}},
  \href{https://doi.org/10.1103/PhysRevD.107.016012}{\emph{Phys. Rev. D}
  {\bfseries 107} (2023) 016012}
  [\href{https://arxiv.org/abs/2209.13441}{{\ttfamily 2209.13441}}].

\bibitem{Ruzi:2024cbt}
A.~Ruzi, Y.~Wu, R.~Ding, S.~Qian, A.M.~Levin and Q.~Li, \emph{{Testing Bell
  inequalities and probing quantum entanglement at a muon collider}},
  \href{https://doi.org/10.1007/JHEP10(2024)211}{\emph{JHEP} {\bfseries 10}
  (2024) 211} [\href{https://arxiv.org/abs/2408.05429}{{\ttfamily
  2408.05429}}].

\bibitem{Aguilar-Saavedra:2024whi}
J.A.~Aguilar-Saavedra, \emph{{Tripartite entanglement in
  H\textrightarrow{}ZZ,WW decays}},
  \href{https://doi.org/10.1103/PhysRevD.109.113004}{\emph{Phys. Rev. D}
  {\bfseries 109} (2024) 113004}
  [\href{https://arxiv.org/abs/2403.13942}{{\ttfamily 2403.13942}}].

\bibitem{Kauffman:1996ix}
R.P.~Kauffman, S.V.~Desai and D.~Risal, \emph{{Production of a Higgs boson plus
  two jets in hadronic collisions}},
  \href{https://doi.org/10.1103/PhysRevD.58.119901}{\emph{Phys. Rev. D}
  {\bfseries 55} (1997) 4005}
  [\href{https://arxiv.org/abs/hep-ph/9610541}{{\ttfamily hep-ph/9610541}}].

\bibitem{Georgi:1977gs}
H.M.~Georgi, S.L.~Glashow, M.E.~Machacek and D.V.~Nanopoulos, \emph{{Higgs
  Bosons from Two Gluon Annihilation in Proton Proton Collisions}},
  \href{https://doi.org/10.1103/PhysRevLett.40.692}{\emph{Phys. Rev. Lett.}
  {\bfseries 40} (1978) 692}.

\bibitem{Zagoskin:2015sca}
T.V.~Zagoskin and A.Y.~Korchin, \emph{{Decays of a neutral particle with zero
  spin and arbitrary CP parity into two off-mass-shell Z bosons}},
  \href{https://doi.org/10.1134/S1063776116020229}{\emph{J. Exp. Theor. Phys.}
  {\bfseries 122} (2016) 663}
  [\href{https://arxiv.org/abs/1504.07187}{{\ttfamily 1504.07187}}].

\bibitem{Godbole:2007cn}
R.M.~Godbole, D.J.~Miller and M.M.~Muhlleitner, \emph{{Aspects of CP violation
  in the H ZZ coupling at the LHC}},
  \href{https://doi.org/10.1088/1126-6708/2007/12/031}{\emph{JHEP} {\bfseries
  12} (2007) 031} [\href{https://arxiv.org/abs/0708.0458}{{\ttfamily
  0708.0458}}].

\bibitem{Bolognesi:2012mm}
S.~Bolognesi, Y.~Gao, A.V.~Gritsan, K.~Melnikov, M.~Schulze, N.V.~Tran et~al.,
  \emph{{On the Spin and Parity of a Single-Produced Resonance at the LHC}},
  \href{https://doi.org/10.1103/PhysRevD.86.095031}{\emph{Phys. Rev. D}
  {\bfseries 86} (2012) 095031}
  [\href{https://arxiv.org/abs/1208.4018}{{\ttfamily 1208.4018}}].

\bibitem{Hagiwara:1993ck}
K.~Hagiwara, S.~Ishihara, R.~Szalapski and D.~Zeppenfeld, \emph{{Low-energy
  effects of new interactions in the electroweak boson sector}},
  \href{https://doi.org/10.1103/PhysRevD.48.2182}{\emph{Phys. Rev. D}
  {\bfseries 48} (1993) 2182}.

\bibitem{Christensen:2008py}
N.D.~Christensen and C.~Duhr, \emph{{FeynRules - Feynman rules made easy}},
  \href{https://doi.org/10.1016/j.cpc.2009.02.018}{\emph{Comput. Phys. Commun.}
  {\bfseries 180} (2009) 1614}
  [\href{https://arxiv.org/abs/0806.4194}{{\ttfamily 0806.4194}}].

\bibitem{Alloul:2013bka}
A.~Alloul, N.D.~Christensen, C.~Degrande, C.~Duhr and B.~Fuks, \emph{{FeynRules
  2.0 - A complete toolbox for tree-level phenomenology}},
  \href{https://doi.org/10.1016/j.cpc.2014.04.012}{\emph{Comput. Phys. Commun.}
  {\bfseries 185} (2014) 2250}
  [\href{https://arxiv.org/abs/1310.1921}{{\ttfamily 1310.1921}}].

\bibitem{Degrande:2011ua}
C.~Degrande, C.~Duhr, B.~Fuks, D.~Grellscheid, O.~Mattelaer and T.~Reiter,
  \emph{{UFO - The Universal FeynRules Output}},
  \href{https://doi.org/10.1016/j.cpc.2012.01.022}{\emph{Comput. Phys. Commun.}
  {\bfseries 183} (2012) 1201}
  [\href{https://arxiv.org/abs/1108.2040}{{\ttfamily 1108.2040}}].

\bibitem{Darme:2023jdn}
L.~Darm\'e et~al., \emph{{UFO 2.0: the \textquoteleft{}Universal Feynman
  Output\textquoteright{} format}},
  \href{https://doi.org/10.1140/epjc/s10052-023-11780-9}{\emph{Eur. Phys. J. C}
  {\bfseries 83} (2023) 631}
  [\href{https://arxiv.org/abs/2304.09883}{{\ttfamily 2304.09883}}].

\bibitem{Alwall:2014hca}
J.~Alwall, R.~Frederix, S.~Frixione, V.~Hirschi, F.~Maltoni, O.~Mattelaer
  et~al., \emph{{The automated computation of tree-level and next-to-leading
  order differential cross sections, and their matching to parton shower
  simulations}}, \href{https://doi.org/10.1007/JHEP07(2014)079}{\emph{JHEP}
  {\bfseries 07} (2014) 079} [\href{https://arxiv.org/abs/1405.0301}{{\ttfamily
  1405.0301}}].

\bibitem{Rahaman:2021fcz}
R.~Rahaman and R.K.~Singh, \emph{{Breaking down the entire spectrum of spin
  correlations of a pair of particles involving fermions and gauge bosons}},
  \href{https://doi.org/10.1016/j.nuclphysb.2022.115984}{\emph{Nucl. Phys. B}
  {\bfseries 984} (2022) 115984}
  [\href{https://arxiv.org/abs/2109.09345}{{\ttfamily 2109.09345}}].

\bibitem{Boudjema:2009fz}
F.~Boudjema and R.K.~Singh, \emph{{A Model independent spin analysis of
  fundamental particles using azimuthal asymmetries}},
  \href{https://doi.org/10.1088/1126-6708/2009/07/028}{\emph{JHEP} {\bfseries
  07} (2009) 028} [\href{https://arxiv.org/abs/0903.4705}{{\ttfamily
  0903.4705}}].

\bibitem{Horodecki:2009ZZ}
R.~Horodecki, P.~Horodecki, M.~Horodecki and K.~Horodecki, \emph{{Quantum
  entanglement}}, \href{https://doi.org/10.1103/RevModPhys.81.865}{\emph{Rev.
  Mod. Phys.} {\bfseries 81} (2009) 865}
  [\href{https://arxiv.org/abs/quant-ph/0702225}{{\ttfamily
  quant-ph/0702225}}].

\bibitem{Yu_2021}
Y.~Yu, \emph{Advancements in applications of quantum entanglement},
  \href{https://doi.org/10.1088/1742-6596/2012/1/012113}{\emph{Journal of
  Physics: Conference Series} {\bfseries 2012} (2021) 012113}.

\bibitem{bennett1996mixed}
C.H.~Bennett, D.P.~DiVincenzo, J.A.~Smolin and W.K.~Wootters, \emph{Mixed-state
  entanglement and quantum error correction}, {\emph{Physical Review A}
  {\bfseries 54} (1996) 3824}.

\bibitem{wootters1998entanglement}
W.K.~Wootters, \emph{Entanglement of formation of an arbitrary state of two
  qubits}, {\emph{Physical Review Letters} {\bfseries 80} (1998) 2245}.

\bibitem{vedral1997quantifying}
V.~Vedral, M.B.~Plenio, M.A.~Rippin and P.L.~Knight, \emph{Quantifying
  entanglement}, {\emph{Physical Review Letters} {\bfseries 78} (1997) 2275}.

\bibitem{wei2003geometric}
T.-C.~Wei and P.M.~Goldbart, \emph{Geometric measure of entanglement and
  applications to bipartite and multipartite quantum states}, {\emph{Physical
  Review A} {\bfseries 68} (2003) 042307}.

\bibitem{vidal2002computable}
G.~Vidal and R.F.~Werner, \emph{Computable measure of entanglement},
  {\emph{Physical Review A} {\bfseries 65} (2002) 032314}.

\bibitem{christandl2004squashed}
M.~Christandl and A.~Winter, \emph{“squashed entanglement”: an additive
  entanglement measure}, {\emph{Journal of mathematical physics} {\bfseries 45}
  (2004) 829}.

\bibitem{yang2009squashed}
D.~Yang, K.~Horodecki, M.~Horodecki, P.~Horodecki, J.~Oppenheim and W.~Song,
  \emph{Squashed entanglement for multipartite states and entanglement measures
  based on the mixed convex roof}, {\emph{IEEE Transactions on Information
  Theory} {\bfseries 55} (2009) 3375}.

\bibitem{hayden2001asymptotic}
P.M.~Hayden, M.~Horodecki and B.M.~Terhal, \emph{The asymptotic entanglement
  cost of preparing a quantum state}, {\emph{Journal of Physics A: Mathematical
  and General} {\bfseries 34} (2001) 6891}.

\bibitem{Song_2016}
W.~Song, L.~Chen and Z.-L.~Cao, \emph{Lower and upper bounds for entanglement
  of renyi-$\alpha$ entropy},
  \href{https://doi.org/10.1038/s41598-016-0029-9}{\emph{Scientific Reports}
  {\bfseries 6} (2016) }.

\bibitem{chen2005concurrence}
K.~Chen, S.~Albeverio and S.-M.~Fei, \emph{Concurrence of arbitrary dimensional
  bipartite quantum states}, {\emph{Physical review letters} {\bfseries 95}
  (2005) 040504}.

\bibitem{mintert2007observable}
F.~Mintert and A.~Buchleitner, \emph{Observable entanglement measure for mixed
  quantum states}, {\emph{Physical review letters} {\bfseries 98} (2007)
  140505}.

\bibitem{zhang2008observable}
C.-J.~Zhang, Y.-X.~Gong, Y.-S.~Zhang and G.-C.~Guo, \emph{Observable estimation
  of entanglement for arbitrary finite-dimensional mixed states},
  {\emph{Physical Review A—Atomic, Molecular, and Optical Physics} {\bfseries
  78} (2008) 042308}.

\bibitem{Zhu_2012}
X.-N.~Zhu and S.-M.~Fei, \emph{Improved lower and upper bounds for entanglement
  of formation},
  \href{https://doi.org/10.1103/physreva.86.054301}{\emph{Physical Review A}
  {\bfseries 86} (2012) }.

\bibitem{Ashby-Pickering:2022umy}
R.~Ashby-Pickering, A.J.~Barr and A.~Wierzchucka, \emph{{Quantum state
  tomography, entanglement detection and Bell violation prospects in weak
  decays of massive particles}},
  \href{https://doi.org/10.1007/JHEP05(2023)020}{\emph{JHEP} {\bfseries 05}
  (2023) 020} [\href{https://arxiv.org/abs/2209.13990}{{\ttfamily
  2209.13990}}].

\end{thebibliography}\endgroup

\end{document}